\documentclass[aps,twocolumn,floatfix,amsmath,amssymb,superscriptaddress]{revtex4-1}
\usepackage{amsmath}
\usepackage{graphicx}
\usepackage{units, siunitx}
\usepackage{braket}

\usepackage{hyperref}
\hypersetup{colorlinks=true, urlcolor=blue, linkcolor=blue, citecolor=blue}

\usepackage{color}
\definecolor{purple}{rgb}{0.5,0,0.5}

\begin{document}

\title{1D Error Correcting Code for Majorana Qubits}

\author{John P. T. Stenger}
\affiliation{Department of Physics and Astronomy, University of Pittsburgh, Pittsburgh, PA, 15260, USA}
\affiliation{Pittsburgh Quantum Institute, Pittsburgh, PA, 15260, USA}
\author{Roger S. K. Mong}
\affiliation{Department of Physics and Astronomy, University of Pittsburgh, Pittsburgh, PA, 15260, USA}
\affiliation{Pittsburgh Quantum Institute, Pittsburgh, PA, 15260, USA}

\begin{abstract}
Although Majorana platforms are promising avenues to realizing topological quantum computing, they are still susceptible to errors from thermal noise and other sources.
We show that the error rate of Majorana qubits can be drastically reduced using a 1D repetition code.
The success of the code is due the imbalance between the phase error rate and the flip error rate.
We demonstrate how a repetition code can be naturally constructed from segments of Majorana nanowires.  We find the optimal lifetime may be extended from a millisecond to over one second.
\end{abstract}

\maketitle

%\section{Introduction}
The main road block in achieving quantum computation~\cite{Galindo2002} is dealing with quantum error.  Isolating a bit of quantum information from its environment is challenging enough, however, in order to realize a useful quantum computation machine it is necessary to maintain coherence for thousands of entangled qubits.
Topological qubits are useful in that they have built-in fault tolerance due to the spatial separations between the anyons and the boundary modes~\cite{Kitaev2003}.  Majorana zero modes~\cite{Volovik:ZeroModes:87-En,Kitaev2001,Read2000}, which appear as end modes of $p$-wave superconducting nanowires, are one the most promising directions in topological quantum computing~\cite{Kitaev2001, Nayak2008, Oreg2010, Lutchyn2010, Alicea2010, Sau2010, Beenakker2013, Mourik2012, Chen2017, Deng2016}.
These Majorana end modes can store information non-locally and can be braided to perform topologically protected logic gates~\cite{Alicea2011, Halperin2012, Hyart2013, Heck2012, Hassler2010, Bonderson2009, Vijay2016, Stenger2019}.  

Although topological qubits have some level of protection from error they will still require error correction in order to be fully implemented as computational qubits.  A perfect Majorana qubit would be infinitely long and held at zero temperature.  Nonzero temperatures lead to a finite quasiparticle density, which will cause errors in the qubit.  There exist error correction codes such as the toric code~\cite{Kitaev2003}, surface codes~\cite{Dennis2002, Freedman2001, Bravyi1998, Fowler2012}, and color codes~\cite{Bombin2006, Landahl2011,Nigg2014}, which can be implemented on Majorana qubits~\cite{Bravyi2010,Vijay2015,Karzig2017, Litinski2017, Li2018, Litinski2018, Vijuela2019, Plugge2016} or in other schemes such as planar codes~\cite{Bombin2007, Bombin2009}.  However, these error correction schemes require a great deal of overhead, having a large number of redundant qubits in order to catch and correct error.  As Kitaev pointed out~\cite{Kitaev2003}, any topological phase of matter can be identified as an error correcting code.  In this vain, we ask if the 1D fermionic topological phase~\cite{Turner2011,Fidkowski2011} built from a chain of Majorana nanowires can be identified with a ``fermion-parity protected error correcting code''.  Provided that fermion parity is conserved, such a chain would protect against quantum errors and would require only a line of physical qubits instead of a surface.
%As we will show, the chain by itself only protects bit-flip error, however, phase error can be corrected using the 1D repetition code.  

%, and the hyperbolic versions of these codes~\cite{Albuquerque2009, Silva2018}

In this paper, we show how a chain of Majorana nanowires can be used to significantly improve the qubit lifetime, because of a hierarchy of different error types in Majorana qubits.  Due to an unexpectedly high observed density of quasiparticles~\cite{Pop2014, Higginbotham2015,Martinis2009,Visser2011, Saira2012, Riste2013}, we argue that phase errors in Majorana qubits are orders of magnitude greater than bit flip errors.  This phase error can be corrected at the expense of the much smaller bit flip error using the repetition code~\cite{Shor:Code:95, Aliferis2008, DevittMunroNemoto:QECReview:2013}.  We describe the repetition code in the language of Majorana qubits.  The code works on a chain of several qubits by measuring the local parity of the chain links.  The simplicity of the code will likely make it experimentally practical in the very near future.  In fact, the repetition code has already been realized for chains of transmon qubits~\cite{Kelly2015}.  However, the repetition code is particularly advantageous for Majorana qubits because of the imbalance of error rates.

We argue that the largest contributing error is phase error, which we estimate from Refs.~\onlinecite{Knapp2018,Knapp2018b,Higginbotham2015,Pop2014,Albrecht2016,Plugge2017} to be in the milliseconds regime.
We find that nine nanowire segments are enough to do efficient error correction and that the optimal length of each segment is about five microns.  For these parameters, we find an improved Majorana qubit lifetime to be on the order of one second.

%\section{Model}
%%%%%%%%%%%%%%%%%%%%%%%%%%%%%%%%
%%%%%%%%%%%%%%%%%%%%%%%%%%%%
\begin{figure}[t]
\includegraphics[width=\columnwidth]{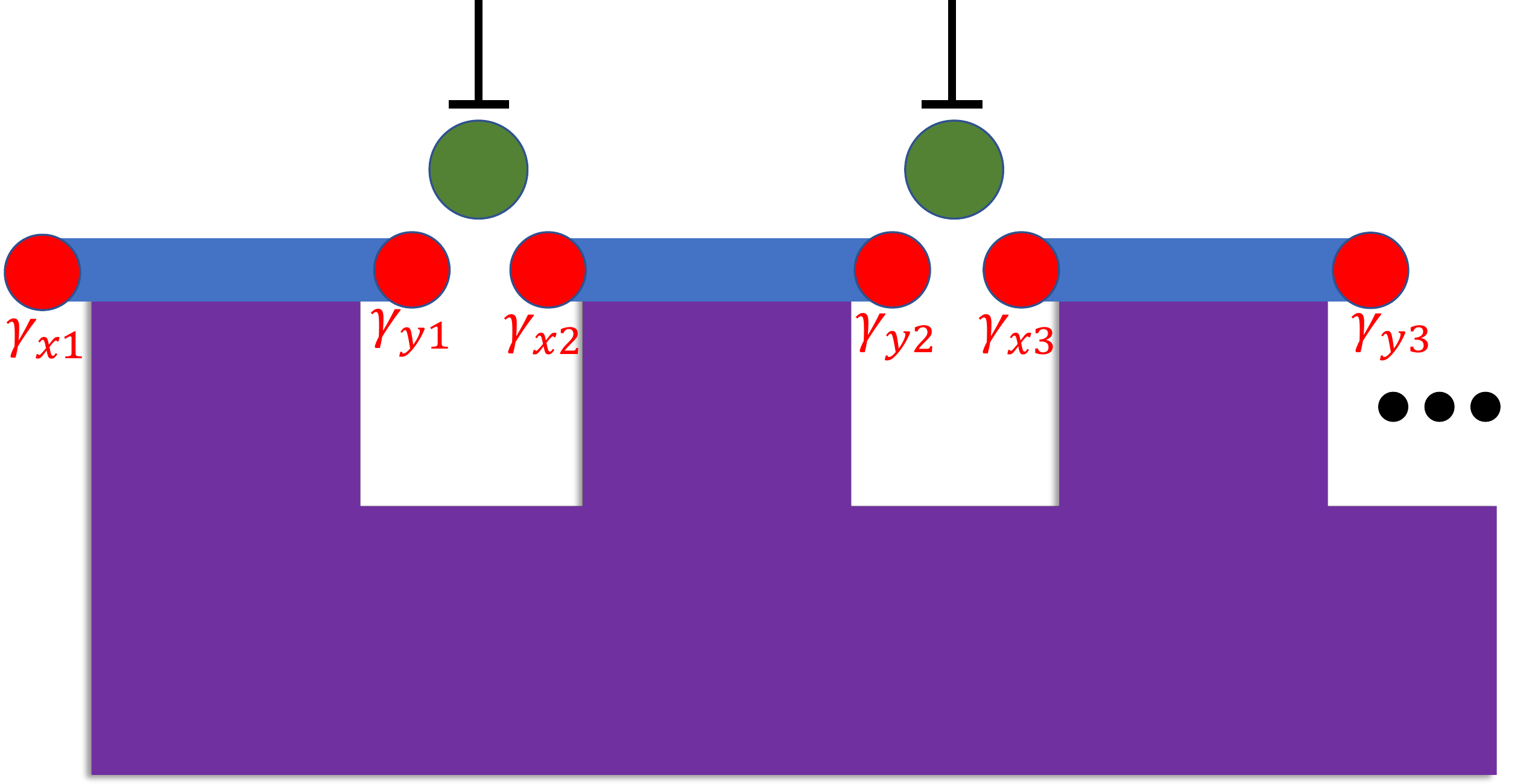}

\caption{Three semiconducting nanowires (blue) coupled to an s-wave superconductor (purple) and driven into the topologically nontrivial phase so that each nanowire has Majorana end modes (red).  The leftmost and rightmost Majoranas are used to store the logical qubit.  The intermediate Majoranas are coupled to a measurement device which is depicted as quantum dots but any measurement device is valid.}
\label{Fig_0}
\end{figure}
%%%%%%%%%%%%%%%%%%%%%%	
%%%%%%%%%%%%%%%%%%%%%%%%%%%%%%%%

The hardware for the Majorana repetition code is composed of segments of topological superconducting nanowires as depicted in Fig.~\ref{Fig_0}.  In the figure, we have $N=3$ nanowire segments but in general we consider arbitrary numbers of segments.  Each segment hosts two Majorana bound states which form a qubit.  (This is in contrast to the definite-parity Majorana qubits which are composed of four Majoranas~\cite{Karzig2017, Litinski2017}.)
%Fermion parity is not conserved in each wire because all the nanowires are connected via the parent superconductor which quenches the charging energy.  For this reason, we will refer to each segment as a physical qubit and calculate the lifetime of each qubit.  
The Hamiltonian for the device is
\begin{align}
    \label{Hv1}
    H=i\Lambda \sum_{i=1}^N\gamma_{xi}\gamma_{yi}+i\sum_{i=1}^{N-1}g_i\gamma_{yi}\gamma_{x(i+1)} ,
\end{align}
where $\Lambda$ decreases as $e^{-L/\xi}$ and which we assume to be homogeneous to each segment, $L$ is the length of each segment, $\xi$ is the Majorana decay length, and $g_i$ is the coupling between Majoranas at the junction between segments $i$ and $i+1$.
There are some external control elements that can be used to tune $g_i$ and to measure the parity ($i\gamma_{yi}\gamma_{x(i+1)}$) of the inner Majoranas.  In Fig.~\ref{Fig_0} these control elements are depicted as quantum dots, however, any measurement technique could be used~\cite{Vijay2016,Grimsmo2019}). 

To analyze errors on each qubit, we couple the system to a phonon bath as in Ref.~\onlinecite{Knapp2018}.  This interaction can cause errors in the occupation of each qubit by promoting electrons into the gap creating Bogoliubov de Gennes particles (i.e., quasiparticles).  Since Refs.~\onlinecite{Pop2014, Higginbotham2015,Martinis2009,Visser2011, Saira2012, Riste2013} measures quasiparticle densities which are much larger than expected at thermal equilibrium, we attribute these observations to either non-thermal quasiparticles in the bulk, or quasiparticles pinned to surface defects.
We therefore assume that these quasiparticles have low enough mobility orthogonal to the nanowire that the excitation stays in the vicinity of the nanowire but high enough mobility along the nanowire that they can travel from one side of the nanowire to the other.  We will discuss two distinct types of qubit errors: flip error in which the parity of a single qubit changes, and phase error in which the two parity states of a qubit incur a relative phase.
%\john{check this}
%\begin{equation}
%H_{\theta}=\int\frac{d^3q}{(2\pi)^3}\sum_{k,i} m_k(-q)(c^{\dagger}_{i,k}+c_{i,k})(\gamma_{xi}+\gamma_{yi})\theta(q) 
%\label{Htheta}
%\end{equation}
%where $\theta(q)$ is a phonon operator, $c_{i,k}$ and $c_{i,k}^{\dagger}$ are the excited state electron operators in segment $i$, and $m_k(-q)$ is the matrix element between excited quasiparticles and Majorana zero modes. This interaction can cause errors in the occupation of each qubit by promoting electrons into the excited state.  We will discuss two distinct types of qubit errors: flip error in which the parity of a single qubit changes and phase error in which the two parity states of a qubit pick up a different phase.

%\section{Error Rates}
%At zero temperature, hybridization of the Majorana end modes is exponentially (as a function of the separation of the modes) suppressed.  Quasi-particle poisoning is protected instead by the superconducting gap of the device. 
It has been argued that the main source of quasiparticle poisoning is mediated by the electron-phonon interaction~\cite{Knapp2018,Aseev2019,Goldstein2011}.   Theoretically, the rate that phonons split apart Cooper pairs and one of the electrons from the pair changes the occupation of the Majorana mode goes as $\Gamma_{\Delta}=\tau_0^{-1}\exp[-\Delta\beta]$ where $\tau_0$ is the characteristic timescale describing electron-phonon coupling, $\Delta$ is the superconducting gap and $\beta$ is the inverse temperature.  For bulk InAs the electron-phonon coupling timescale is on the order of tens of nanoseconds ($\tau_0\approx \unit[10]{ns}$) \cite{Knapp2018}.

The exponentially decaying characteristic of the Cooper pair breaking rate saturates at low temperature where relatively large non-thermal quasiparticle densities have been observed experimentally~\cite{Pop2014, Higginbotham2015,Martinis2009,Visser2011, Saira2012, Riste2013}.  In this case the exponential is replaced by the quasi-particle density in the following way~\cite{Knapp2018}:
\begin{align}
    \label{rate0}
    \Gamma_{\Delta}=\tau_0^{-1}\sqrt{2\Delta\beta/\pi} \, n_{qp}V
\end{align}
where $n_{qp}$ is the quasi-particle density and $V$ is the volume of the superconductor in the vicinity of the edge modes. For typical Al coated InAs devices at temperatures of $\Delta\beta\approx 10$ \cite{Knapp2018}, the factor under the square root is order unity.  We take $\Delta\approx\unit[0.2]{meV}$~\cite{Knapp2018, Das2012, Higginbotham2015} to be the Al gap.  The relevant volume depends on the decay rate of the Majorana modes $\xi\approx\SI{0.1}{\micro m}$~\cite{Albrecht2016} which gives a volume of $V=\xi^3$.
%coherence length $\xi=\nu \hbar/\pi \Delta$ of the Al superconductor.  Using the femi-velocity for Al ($\nu\approx \unit[10^5]{m/s}$)~\cite{Knapp2018}, we find the relevant volume to be $V=\xi^3=\unit[10^{-3}]{\mu m}$.  
%The total volume of a typical Al shell is $V_{total}\approx10^{-4}~\si{\micro\meter^3}$~\cite{Higginbotham2015}, however, the relevant volume could be reduced by about a factor of ten; $V=(\xi/L) V_{total}\approx \unit[10^{-5}]{\mu m^3}$, where $L$ is the length of the superconductor~\cite{Knapp2018}. 
There is a large range of experimental data for the non-thermal quasi-particle density in Al, $n_{qp}\approx 0.01$ to $\SI{10}{\micro m^{-3}}$~\cite{Pop2014, Higginbotham2015,Martinis2009,Visser2011, Saira2012, Riste2013}, which results in a range of timescales $1/\Gamma_{\Delta}\approx \unit[1]{ms}$ to $\SI{1}{\micro s}$ which are all a significant source of error for quantum computation.  

The reverse process of a pair of electrons recombining into a Cooper pair is much faster $\Gamma_{0}=\tau_0^{-1}$ which means a stable flip in the occupation state is rare.  However, recombination is equally likely to occur in either Majorana mode giving a fifty percent chance of a phase error upon recombination~\cite{Knapp2018b}.    

We can calculate the probability of flip errors and the probability of phase errors by considering a length of nanowire with two Majorana end modes ($\gamma_x$ and $\gamma_y$) which are occupied by an electron.  Assume that the electron is about to interact with a phonon and get promoted to the superconducting gap through $\gamma_x$.  Then we can ask what is the probability $P_{f}$ that it stays at the gap, what is the probability $P_{\phi}$ that it returns through $\gamma_y$, and what is the probability $P_0$ that it either does not get excited or returns through $\gamma_x$.  Then $P_f$ is the flip error probability and $P_{\phi}$ is the phase error probability.  The rate equation is given as,
\begin{align}
    \begin{bmatrix} \dot{P}_{0} \\ \dot{P}_{f} \\ \dot{P}_{\phi} \end{bmatrix}
    &=
    \begin{bmatrix}
    -\Gamma_{\Delta} && \Gamma_{0} && 0
    \\
    \Gamma_{\Delta} && -2\Gamma_{0} && \Gamma_{\Delta}
    \\
    0 && \Gamma_{0} && -\Gamma_{\Delta}
    \end{bmatrix}
    \begin{bmatrix}
    P_0 \\ P_f \\ P_{\phi}
    \end{bmatrix},
\end{align}
using the initial condition $\big(P_0(0),P_f(0),P_{\phi}(0)\big) = \big(1,0,0\big)$, the solution is
\begin{align}\begin{split}
    P_0(t)&=\frac{2\Gamma_{0}(1+e^{-\Gamma_{\Delta}t})+\Gamma_{\Delta}(1+e^{-2\Gamma_{0}t-\Gamma_{\Delta}t})}{4\Gamma_{0}+2\Gamma_{\Delta}} ,
    \\
    P_{f}(t)&=\frac{(1-e^{-2\Gamma_{0}t-\Gamma_{\Delta}t})\Gamma_{\Delta}}{2\Gamma_{0}+\Gamma_{\Delta}} ,
    \\
    P_{\phi}(t)&=\frac{2\Gamma_{0}(1-e^{-\Gamma_{\Delta}t})+\Gamma_{\Delta}(-1+e^{-2\Gamma_{0}t})e^{-\Gamma_{\Delta }t}}{4\Gamma_{0}+2\Gamma_{\Delta}} .
\label{eq:probs}
\end{split}\end{align}

While the phase error approaches $\approx 1/2$ at large time, the flip error saturates to $P_{f}(\infty)\approx\Gamma_{\Delta}/(2\Gamma_{0})$.  Using the parameters defined in this section, we have a range of $P_{f}(\infty)\approx 5\times10^{-3}$ to $5\times10^{-6}$ depending on the quasiparticle density.  At a measurement time of $t=\SI{100}{\micro s}$, the ratio of the two errors ranges from $P_{\phi}(t)/P_f(t)\approx 10$ to $10^5$.
The large discrepancy between these error probabilities allows us to correct the phase error at the expense of the flip error using the repetition code. 

We note that our analysis is valid for short to intermediate times.  On the long time scale, mobile quasiparticles (in the bulk superconductor) can carry fermions away from the Majorana modes increasing the probability of a flip error.  Our model is justified by assuming that the majority of quasiparticles originates from the nanowire-superconductor interface and are localized.

We also include the effect of Majorana hybridization error via Hamiltonian evolution.
The Hamiltonian is written in Eq.~\eqref{Hv1} where we assume that $g_i\ll\Lambda$ between measurements.  The time scale for Hamiltonian evolution is $\hbar/\Lambda$ which becomes comparable to the phase error timescale for $\Lambda\approx 0.1~\si{\pico\electronvolt}$, (where we have taken the low end of the quasiparticle density range).  The energy $\Lambda$ corresponds to a length from $\Lambda=\Lambda_0 e^{-L/\xi}$, where  $\Lambda_0\approx\unit[0.3]{meV}$ and the decay length $\xi\approx\unit[260]{nm}$~\cite{Albrecht2016}.
%At a length of $L\approx\SI{5}{\micro m}$ the hybridization timescale gets below the phase error timescale.
Therefore, if we set the length of each segment of nanowire to be greater than $L\approx\SI{5}{\micro m}$, then the quasiparticle induced phase error is dominant over that from hybridization.

%\section{The Repetition Code}

\begin{figure}[t]
    \includegraphics[width=0.4\textwidth]{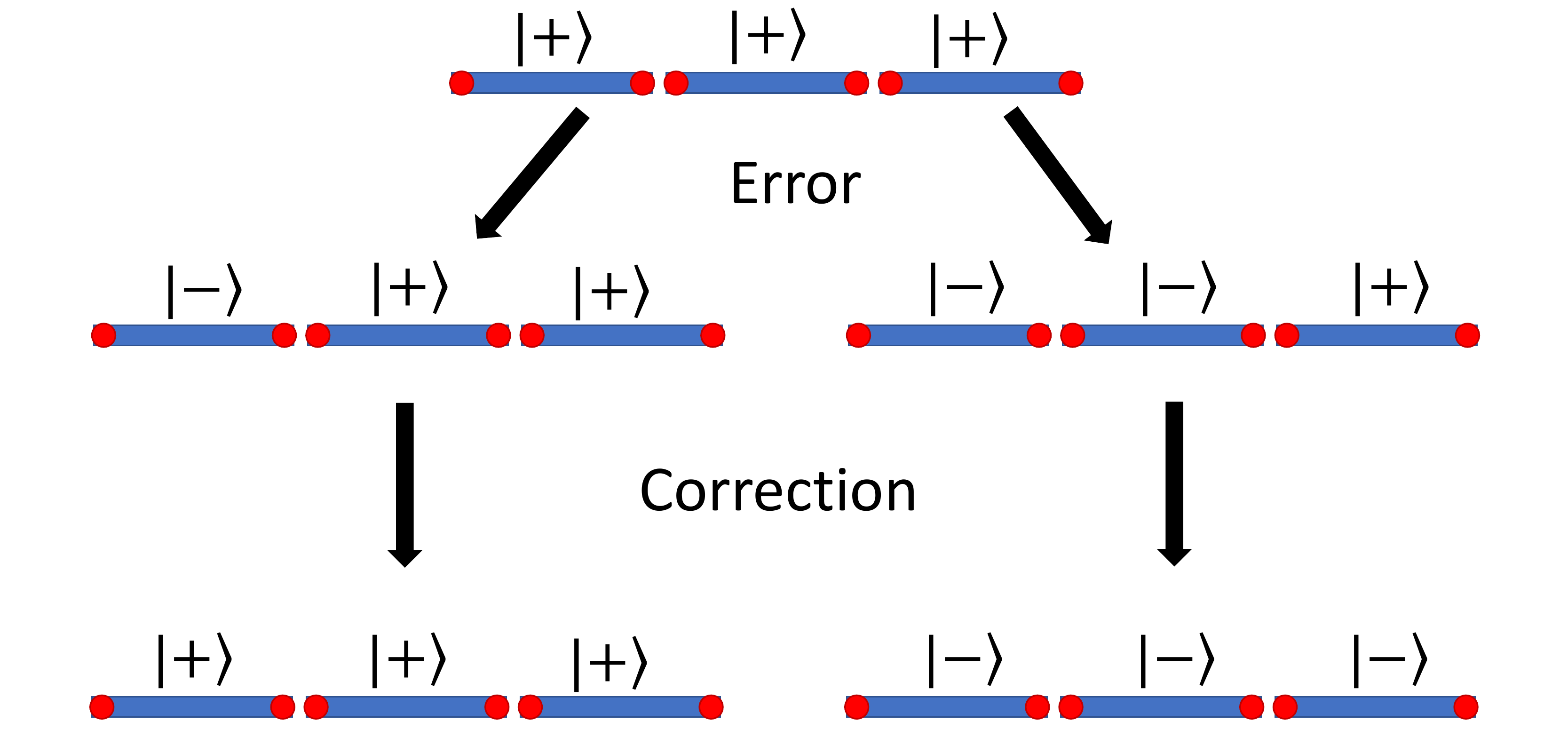}
    \caption{Depiction of the error correction process.  The logical qubit is represented by the state $\ket{+++}$ which is composed of the physical qubits depicted in the top row.  Error can cause a number of the physical qubits to change as depicted in the second row.  The error correction process brings the system back into the logical basis, as depicted in the bottom row.  This may restore the original state (left), but may also project onto the wrong state (right) if too much error has occurred.}
    \label{Fig_1}
\end{figure}

The standard repetition code~\cite{DevittMunroNemoto:QECReview:2013} works as follows.  Suppose we want to encode a physical qubit $\ket{\psi}$.  First write the state in the basis of eigenstates of $X = \sigma^x$: i.e., $\ket{\psi}=a\ket{+}+b\ket{-}$.   We encode the logical qubit into multiple ($N=3$) physical qubits via $\ket{\psi}\to a\ket{+++}+b\ket{---}$.  Since the logical basis ($\ket{+++}$ and $\ket{---}$) is now smaller than the total space of states, we can repeatedly measure certain syndromes of each physical qubit and catch errors without destroying the logical qubit (see Fig.~\ref{Fig_1}).  The syndromes are operators for which the logical basis state are degenerate eigenstates, such as $X_i X_j$.
A phase error swaps $\ket{+}\rightleftarrows\ket{-}$ (i.e., applies $Z=\sigma^z$ to one of the physical qubit), such error takes the state out of the logical basis.  As long as only a single qubit has incurred a $Z$-error, it can be detected and corrected by undoing the error.  (See table~\ref{tab:T1}.)

If an error occurred in more than one qubit then we would accidentally project onto the wrong logical basis.  However, we can always improve the amount of acceptable error by increasing the number of physical qubits $N$ in the encoding.  On the other hand, the repetition code does not prevent flip errors, i.e., $\ket{\pm}\rightarrow\pm\ket{\pm}$.
(A repetition code in the $Z$ basis will be able to correct flip $X$-errors at the expense of phase $Z$-errors no longer being correctable.)

%%%%%%%%%%%%%%%%%%%%%%%%%%%%%%%%
%%%%%%%%%%%%%%%%%%%%%%%%%%%%
%%%%%%%%%%%%%%%%%%%%%%	
%%%%%%%%%%%%%%%%%%%%%%%%%%%%%%%%

%\section{The Repetition Code in the Language of Majorana Bound States}

\begin{table}
    \renewcommand{\arraystretch}{1.2}
    \begin{tabular}{ c|cc|c| } 
     \qquad\qquad\qquad\qquad & i$\gamma_{y1}\gamma_{x2}$ & i$\gamma_{y2}\gamma_{x3}$ & Error correction \\ 
      & $X_1 X_2$ & $X_2 X_3$ & operation \\\hline
     $\ket{+++}$ & $+$ & $+$ & None \\ 
     $\ket{++-}$ & $+$ & $-$ & $Z_3$\\ 
      $\ket{+-+}$ & $-$ & $-$ & $Z_2$\\ 
      $\ket{-++}$ & $-$ & $+$ & $Z_1$\\ 
      $\ket{+--}$ & $-$ & $+$ & $Z_1$\\ 
      $\ket{-+-}$ & $-$ & $-$ & $Z_2$\\ 
      $\ket{--+}$ & $+$ & $-$ & $Z_3$\\ 
      $\ket{---}$ & $+$ & $+$ & None\\ 
     \hline
    \end{tabular}
    \caption{%
    Error correction for a 3-qubit repetition code.
    The middle columns gives the result of the syndrome measurements $X_iX_{i+1} = i\gamma_{yi}\gamma_{x(i+1)}$.  The last column shows the appropriate correction operator given the results of the syndrome measurement. 
    }
    \label{tab:T1}
\end{table}

The repetition code lends itself naturally to the Majorana system.   The phase error can be understood as a quasiparticle coming into the wire on one end (applying $\gamma_x$) and back out through the other (applying $\gamma_y$).  While a flip error involves only one of $\gamma_x$ or $\gamma_y$.  In order to apply the repetition code, we need at least $2N=6$ Majorana bound states ($\gamma_{x1},\gamma_{y1},\gamma_{x2},\gamma_{y2},\gamma_{x3},\gamma_{y3}$), see Fig.~\ref{Fig_0}.  We can relate Majorana operators to Pauli operators using the Jordan-Wigner transformation: $\gamma_{xi}=X_i\prod_{j<i}Z_j$ and $\gamma_{yi}=-Y_{i}\prod_{j<i}Z_j$.
In this basis, the phase error is simply $Z_{i}=i\gamma_{xi}\gamma_{yi}$.
As before, we can encode the logical qubit into the three (or more) sections of nanowire (i.e., $\ket{+}\rightarrow\ket{+++}$ and $\ket{-}\rightarrow\ket{---}$).
With this encoding, the syndrome operators become bilinears of Majoranas:  $i\gamma_{xi}\gamma_{y(i+1)}=X_iX_{i+1}$.  Table~\ref{tab:T1} shows the results for the three qubit case.
If an error is detected, we can then project back onto the logical space using various parity operators $Z_i=i\gamma_{xi}\gamma_{yi}$.

%\section{Three Qubit Code}

\begin{figure}[t]
\includegraphics[width=\columnwidth]{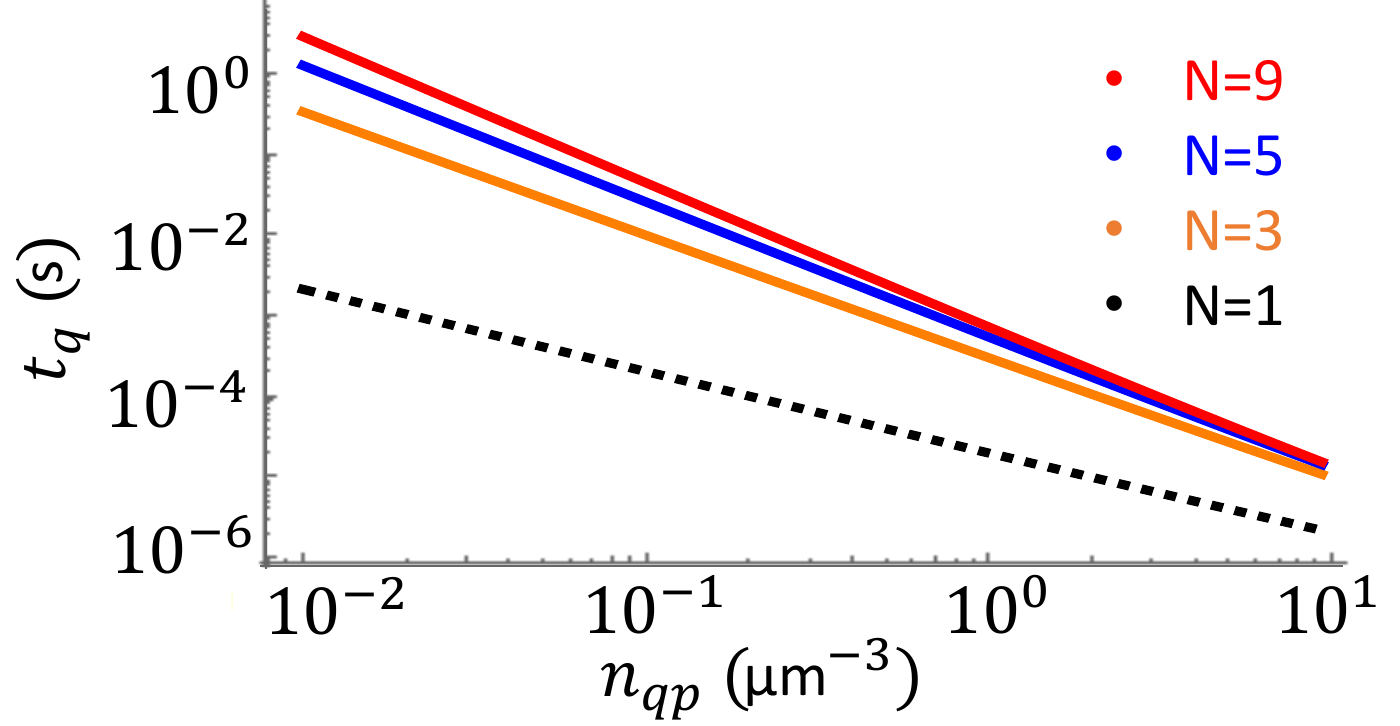}

\caption{Majorana qubit lifetime as a function of quasiparticle density.  The dotted black curve shows the uncorrected lifetime.  The orange, blue, and red curves show the lifetime after error correction for a 3, 5, and 9 qubit code respectively.  The measurement time is optimized given the number of qubits in the code.}
\label{Fig_3}
\end{figure}

We analyze the code using Kraus operators assuming that errors are independent \footnote{see the appendix for details}.  After error correction, the lowest order phase error in the three qubit code goes as $\bar{P}_{\phi}=3P_{\phi}^2+\mathcal{O}(P_{\phi}^3)$~\cite{DevittMunroNemoto:QECReview:2013}, with the first order terms eliminated.
However, the flip error become three times more likely: $\bar{P}_f=3P_f+\mathcal{O}(P_{f}^2)$.  The orange curve in Fig.~\ref{Fig_3} shows the three qubit lifetime after error correction as a function of quasiparticle density.   Applying the three qubit code to the hybridization error we similarly find that the error after correction is $\bar{P}_h=3P_h^2+\mathcal{O}\big(P_h^{3}\big)$ with $P_h=(\Lambda t/\hbar)^2$, again with the first order error removed.
%\textcolor{red}{The similarity to the phase error is no coincidence since the Hamiltonian contains only $Z_i$ operators.  In both cases,}
%We see that the first order error is removed, therefore, even if hybridization is significant we can still do error correction using the repetition code.   

%\section{More Qubits}

Now consider a general (odd) number of Majorana segments $N$.  To leading order, the bit flip error probability after error correction increases $N$-fold:
\begin{equation}
    \bar{P}_f=N P_f+\mathcal{O}(P_{f}^2).
    \label{eq:Pfb}
\end{equation}
%The factor $N$ comes from the fact that each additional qubit provides a channel for a bit flip error $X_i$.
However, $\tfrac{N-1}{2}$ orders of phase error can be removed by error correction.  Therefore, the leading contribution to the error-corrected $\bar{P}_\phi$ must come from $\tfrac{N+1}{2}$ errors:
\begin{align}
    \bar{P}_{\phi}=\binom{N}{\frac{N+1}{2}}P_{\phi}^{\frac{N+1}{2}} + \mathcal{O}\left(P_{\phi}^{\frac{N+3}{2}}\right).
    \label{eq:Pphib}
\end{align}
Similar to the dephasing error, the hybridization error also goes as
\begin{align}
    \bar{P}_h=\binom{N}{\frac{N+1}{2}}P_h^{\frac{N+1}{2}} + \mathcal{O}\left(P_{h}^{\frac{N+3}{2}}\right)
    \label{eq:Phb}
\end{align}
where $P_h=(\Lambda t/\hbar)^2$.

%\section{Discussion}

%\subsection{Optimization}

%%%%%%%%%%%%%%%%%%%%%%%%%%%%%%%%
%%%%%%%%%%%%%%%%%%%%%%%%%%%%
%%%%%%%%%%%%%%%%%%%%%%	
%%%%%%%%%%%%%%%%%%%%%%%%%%%%%%%%

We find the optimal number of qubits $N$ by comparing Eq.~\eqref{eq:Pfb} with \eqref{eq:Pphib}.  We want enough qubits that $\bar{P}_{\phi}$ is pushed down to $\bar{P}_f$.  The optimal amount of time $t$ between measurements depends on $N$.  It has been argued that Majorana qubits can be measured on the $t<\SI{1}{\micro s}$ timescale~\cite{Plugge2017}.  However, since the flip error saturates, it is actually helpful to wait a certain amount of time between error correction implementation.  Using the low end of the reported quasiparticle density ($n_{qp}=\SI{0.01}{\micro m^{-3}}$), we find that error correction begins to saturate around $N=9$ and that the optimal measurement timescale is $t\approx\SI{100}{\micro s}$.  For these parameters, we have $P_{\phi}\approx0.01$, $P_f\approx10^{-5}$, and the probability of an error after error correction is $\bar{P}=\bar{P}_{\phi}=\bar{P}_f\approx10^{-5}$ which means the qubit lifetime is $t_q=t/2\bar{P}\approx\unit[1]{s}$.  

%\john{How important is the rest of this paragraph?} At larger measurement timescales the phase error becomes to large.  For example at $t=\unit[1]{ms}$ the phase error is $P_{\phi}(t)=0.32$ and the amount of qubits needed to make $\bar{P}_{\phi}=\bar{P}_f$ makes $\bar{P}_f$ large and therefore the lifetime is short.  For smaller timescales, the flip error probability decreases much slower than $t$.  For example at $t\approx\SI{10}{\micro s}$ the optimal number of segments is $N=3$ at which point the error corrected probability is still on the order of $\bar{P}\approx10^{-5}$ so that the qubit lifetime is reduced to $t_q=t/2\bar{P}\approx0.1$~seconds.

If we take higher quasiparticle densities, we are still able to improve the lifetime of the qubit.   Figure~\ref{Fig_3} shows both the qubit lifetime as a function of quasiparticle density for a 3 (orange), 5 (green) and 7 (red) qubit code as well as the uncorrected lifetime.    Take, for example, $n_{qp}=\SI{1}{\micro m^{-3}}$.   Applying the repetition code, we find an optimal measurement time of $t=\SI{1}{\micro s}$ and that error correction already begins to saturate at $N=5$.  For these values we have a corrected qubit lifetime of $t_q=t/2\bar{P}=\unit[1]{ms}$.  Although this is far below the qubit lifetime found for small quasiparticle densities, it is still a two orders of magnitude improvement from the bare lifetime which is $t/2P_{\phi}(t)=\SI{5}{\micro s}$ at this quasiparticle density.

%\subsection{Specific Implementation Schemes and Device Integration}
%%%%%%%%%%%%%%%%%%%%%%%%%%%%%%%%
%%%%%%%%%%%%%%%%%%%%%%%%%%%%
\begin{figure}[t]

\includegraphics[width=\columnwidth]{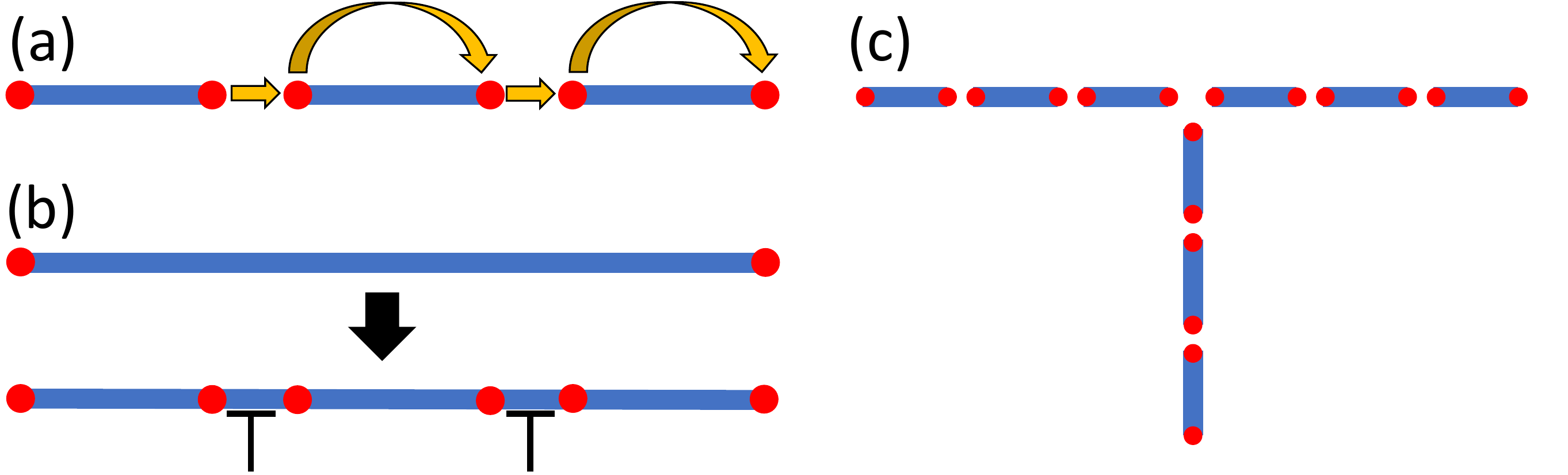}

\caption{Encoding implementations (a and b), and integration of the repetition code into a larger device (c).  (a) Start with three separate nanowire segments and teleport the second Majorana operator to the end.  (b) Start with a single topological superconducting nanowire and divide it into three section using potential gates.    (c) a three segment repetition code integrated into a tri-junction.}
\label{Fig_2}
\end{figure}
%%%%%%%%%%%%%%%%%%%%%%	
%%%%%%%%%%%%%%%%%%%%%%%%%%%%%%%%

Up to this point, we have talked about a generic Majorana repetition code; we now make some comments regarding code implementation.
There are two natural methods of encoding the logical qubit.
If we begin with a state in the leftmost wire segment (in e.g.\ Fig.~\ref{Fig_2}(a)), we can encode the logical qubit by progressively teleporting $\gamma_{y1}$ to the right so that the parity state that was held by the end modes of the first segment is now held by the end modes of the entire wire~\cite{Fu2010, Vijay2016}.  This can be performed by successively measuring syndromes and parity operators.
Alternatively, we could begin by writing the state onto the end modes of the entire wire and then section the wire off using electrostatic contacts, as shown in Fig.~\ref{Fig_2}(b). %\textcolor{red}{Similarly, to decode the state, we can either reverse the teleportation or we can simply measure the state of the furthest end modes.}

There are also two way of dealing with the error recorded by the syndrome measurement.  One method is to correct the error by applying $Z_i$ to the effected physical qubit.  In this way, we catch error as it travels from one end of the wire to the other and send it back.  Another method of dealing with the error is to use the results of the syndrome measurements to update what we consider to be the logical basis~\cite{Knill2005}.  This method is potentially more efficient and reduces the number of gate operations on the system.

Finally, we remark that this error correcting code can be easily integrated into larger Majorana computing schemes by simply replacing single Majorana nanowires by several segments of nanowire.  For example, Fig.~\ref{Fig_2}(c) shows a tri-junction where each arm is composed of three segments and is therefore capable of performing error correction.

%\section{Conclusion}
We have shown that the repetition code can be used to resolve the discrepancy between phase error and bit-flip error in Majorana based qubits.  We find that the lifetime of the qubit can be improved from the ms regime to greater than one second.  Qubit lifetime may be further improved by incorporating the repetition code within other error correction codes.  Therefore, we view the Majorana repetition code as a medium term goal in the broader quest for quantum computation.  Although increasing the separation of Majorana end modes exponentially suppresses hybridization of the modes, our results place a bound on the optimal separation length.  Beyond $L \approx 3\times\SI{5}{\micro m}$, it is more useful to break the nanowire up into multiple segments and perform error correction than to continue increasing the length of the individual segment.  As $\SI{1}{\micro m}$ length nanowires have already been reported~\cite{Albrecht2016}, the community is quickly approach this limit.  It is, therefore, likely that the 1D Majorana code will become practical in the near future, and will act as a stepping stone to fully fault-tolerant topological quantum computing.

\begin{acknowledgements}
We gratefully thank Jason Alicea, Sergey Frolov, Torsten Karzig, Christina Knapp, David Pekker, Falko Pientka, and Felix von~Oppen for helpful discussions.
This work is supported by NSF~PIRE-1743717 and NSF~DMR-1848336.
\end{acknowledgements}

\bibliography{Ref}

\clearpage
\appendix

\renewcommand\thefigure{\thesection.\arabic{figure}}  

\section{Error Model}
In the main text, we discussed the error rates of an $N$ qubit system both before and after error correction.  Here we discuss how those errors are calculated.  We will focus on the $N=3$ case and compare to the $N=1$ (no error correction) case.  The extension to the general $N$ case is straight forward.

We consider three types of error: dephasing, depolarization, and overlapping Majoranas which cause error through Hamiltonian evolution.  As we have argued, the dephasing rate, which depends on $P_{\phi}$, will be much greater than the depolarizing rate, which depends on $P_f$.  Therefore, we prepare the qubit so that error correction can be performed on the dephasing channel.  The initial density operator is given as,
\begin{equation}
\rho_0=\sum_{s,s^{\prime}}\rho_{s,s^{\prime}}\ket{s}\bra{s^{\prime}}
\end{equation}
where $s,s^{\prime}\in\{-,+\}$.

To calculate the error rates we will use the Kraus formalism:
\begin{equation}
\rho(t)=\sum_{i}M_i\rho_0 M_i^{\dagger}
\label{kraus}
\end{equation}
where the Kraus operators $M_i$ are specific to the dissipation channel.  

After we apply the Kraus operators to each qubit we do error correction by applying the measurement projection operators.  We need at least three qubits (six Majoranas) to perform error correction. For the three qubit case, the projection operators are
\begin{align}
    \Pi_{r,s}=\frac{1}{4}\left[1+r(i\gamma_{x1}\gamma_{y2})\right]\left[1+s(i\gamma_{x2}\gamma_{y3})\right].
\end{align}
which are applied to the density matrix.  The corrected density matrix is
\begin{align}
\begin{split}
    \bar{\rho} &= \Pi_{1,1}\rho \Pi_{1,1}+Z_1 \Pi_{-1,1}\rho \Pi_{-1,1}Z_1
    \\
    &\quad+Z_2\Pi_{-1,-1}\rho \Pi_{-1,-1}Z_2+Z_3 \Pi_{1,-1}\rho \Pi_{1,-1}Z_3 ,
    \end{split}
\label{eRcoR}
\end{align}
where $Z_i$ is the Pauli $z$-matrix acting on the $i$\textsuperscript{th} qubit.  We compare the corrected density matrix to the initial one using the fidelity $\bar{P} = 1-F(\rho,\bar\rho) = 1 - \big(\operatorname{Tr}\sqrt{ \sqrt{\rho}\bar{\rho}\sqrt{\rho} }\big)^2$.

\subsection{Dephasing}
The Kraus operators in the dephasing channel are:
\begin{align}\begin{split}
    M_0 &= \sqrt{1-P_{\phi}} ,
    \\
    M_{i,1} & =\sqrt{P_{\phi}}Z_i ,
    \label{phase}
\end{split}\end{align}
where $P_{\phi}$ is the probability of a phase error. Using these in Eq.~\eqref{kraus} we find the time dependent density matrix for a single qubit to be
\begin{equation}
\rho(t)=\sum_{s,s^{\prime}}\left(\rho_{s,s^{\prime}}+P_{\phi}(\rho_{s+1,s^{\prime}+1}-\rho_{s,s^{\prime}})\right)\ket{s}\bra{s^{\prime}}
\end{equation}
where $s$ and $s^{\prime}$ are defined modulo 2, i.e., $(+)+1=(-)$ and $(-)+1=(+)$.

For three qubits, we apply Eq.~\eqref{kraus} once for each.  Then we apply error correction as in Eq~\eqref{eRcoR} to obtain $\bar{\rho}$.  We see that the first order error is removed.  Take for example, $\rho_{++}=1$ and $\rho_{--}=\rho_{+-}=\rho_{-+}=0$ then the fidelity is easy to calculate
\begin{align}
\bar{P}=3P_{\phi}^2-2P_{\phi}^3 .
\label{eq:rhobar}
\end{align}

\subsection{Depolarization}
The Kraus operators in the depolarizing channel are:
\begin{align}\begin{aligned}
    M_0 &= \sqrt{1-\frac{3}{4}P_{pl}} ,
    &
    M_{i,1} &= \sqrt{\frac{1}{4}P_{pl}}X_i ,
    \\
    M_{i,2}&=\sqrt{\frac{1}{4}P_{pl}}Y_i ,
    &
    M_{i,3}&=\sqrt{\frac{1}{4}P_{pl}}Z_i ,
    \label{pol}
\end{aligned}\end{align}
where $P_{pl}$ is the probability of a polarization error.  Using these in Eq.~\eqref{kraus} we find the time dependent density matrix for the single qubit case to be
\begin{align}
    \begin{split}
    \rho(t)=\sum_{s,s^{\prime}}\left(\rho_{s,s^{\prime}}+\frac{1}{2}P_{pl}(\rho_{s+1,s^{\prime}+1}\delta_{s,s^{\prime}}-\rho_{s,s^{\prime}})\right)\ket{s}\bra{s^{\prime}}
    \\
    \quad
    \end{split}
\end{align}
one sees that depolarization causes a mixing of the diagonal terms just like the dephasing.  However, the off diagonal terms are simply damped out.

Because the off diagonal terms damp out, this error is not corrected in the three qubit code.  Take $\rho_{+-}=\rho_{-+}=1/2$ and $\rho_{--}=\rho_{++}=0$ as an example.  In this case,
\begin{align}
\bar{P}=3P_{pl}-3P_{pl}^2+P_{pl}^3 .
\end{align}
The first order error in the off diagonal terms is three times as probable as the single qubit case without error correction.  This off diagonal error is a bit flip (i.e., $X$ type) which, as we have argued, is extremely rare.    

\subsection{Hamiltonian Evolution}
For any finite length of wire, the Majorana modes are not exact eigenstates of the Hamiltonian.  Therefore, Hamiltonian evolution of the system will decohere the qubit.  The time dependent density matrix, in this case, is given by
\begin{align}
    \rho(t)=e^{-i H t/\hbar } \rho_0 e^{i H t/\hbar} ,
    \label{H1site}
\end{align}
where $H=\sum_i\Lambda Z_i$; the coupling $\Lambda$ decays exponentially with the distance between the Majorana modes.  We have dropped the $g_i$ term in Eq.~(1) of the main text because it is assumed to be turned off between measurements.  We find the time dependent density matrix for a single qubit to be
\begin{align}
    \begin{split}
\rho(t)=\frac{1}{2}&\sum_{s,s^{\prime}}
\\
&[\rho_{s,s^{\prime}}(1+\cos(\Lambda t))+\rho_{s+1,s^{\prime}+1}(1-\cos(\Lambda t))
\\
&+i(\rho_{s+1,s^{\prime}}-\rho_{s,s^{\prime}+1})\sin(\Lambda t)]\ket{s}\bra{s^{\prime}} .
    \end{split}
\end{align}
Similar to the dephasing case, hybridization applies $Z_i$ operators to the density matrix.  By applying three qubit error correction we remove the first order error (in $P_h=(\Lambda t)^2$) just like in the phase damping case.  Using $\rho_{++}=1$ and $\rho_{--}=\rho_{+-}=\rho_{-+}=0$ we find that, 
\begin{align}
\bar{P}=3P_h^2+\mathcal{O}(P_h^3) .
\end{align}
This is exactly the same as the lowest order error after applying the repetition code to the dephasing channel.  If we looked at the trace distance instead of the fidelity then the lowest order hybridization error is not identical to the dephasing error.  However, the first order hybridization error is still removed.

\section{Optimizing the qubit lifetime}

%%%%%%%%%%%%%%%%%%%%%%%%%%%%%%%%
%%%%%%%%%%%%%%%%%%%%%%%%%%%%
\begin{figure}[t]
\includegraphics[width=\columnwidth]{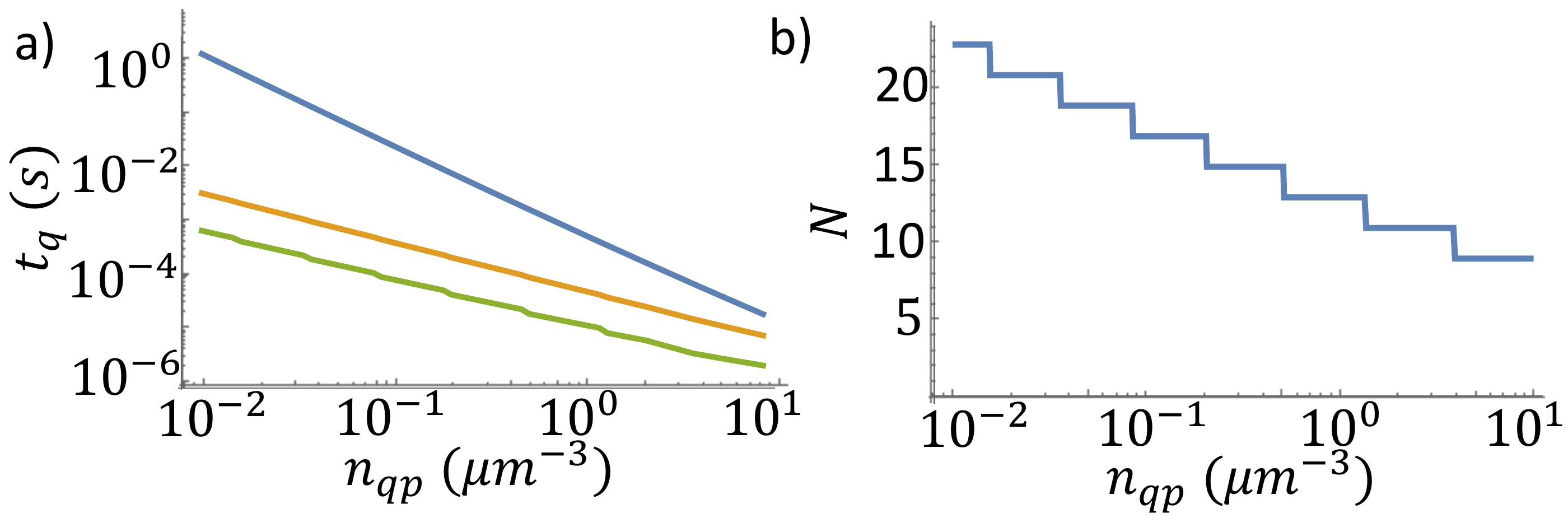}
\caption{Fully optimized qubit lifetimes.  (a) the optimized measurement time $t$ (green), uncorrected qubit lifetime (yellow), and the corrected lifetime $t_q$ (blue) as a function of the quasiparticle density $n_{qp}$.  (b) the optimized number of qubits $N$ as a function of quasiparticle density $n_{qp}$.}
\label{Fig_S1}
\end{figure}
%%%%%%%%%%%%%%%%%%%%%%	
%%%%%%%%%%%%%%%%%%%%%%%%%%%%%%%%

The curves in Fig.~3 of the main text are generated by optimizing the measurement time.  Here we show the optimization procedure in full detail. 

The probabilities of a bit flip error $P_f(t)$ and phase error $P_{\phi}(t)$ without correction are given in Eq.~(4) of the main text and the probabilities after correction $\bar{P}_f(N,t)$, $\bar{P}_{\phi}(N,t)$ are given in terms of $P_f$ and $P_{\phi}$ in Eqs.~(5) and~(6) of the main text.
By selecting a number of qubits $N$, the measurement timescale $t$ can be optimized to result in the greatest qubit lifetime $t_q$ by solving the equation,
\begin{align}
    \bar{P}_f(N,t)=\bar{P}_{\phi}(N,t) .
    \label{barP}
\end{align}
The full equation for qubit lifetime is 
\begin{align}
    t_q=\frac{t}{-\ln{(1-2\bar{P}_f)}}
    =\frac{t}{-\ln{(1-2\bar{P}_{\phi})}} .  
\end{align}
However, for small probabilities we can take $t_q \approx t/2\bar{P}_f=t/2\bar{P}_{\phi}$.  

Alternatively, both the measurement time and number of qubit can be optimized at once.  To do this, we solve Eq.~\eqref{barP} for many values of $N$ and choose the one which makes both probabilities the smallest.  The results of this calculation are shown in Fig.~\ref{Fig_S1}.  We find that the qubit lifetime is not significantly improved from those shown in Fig.~(3) even though the number of qubits is greatly increased at low quasiparticle densities.

\end{document}